\documentclass[letterpaper, 10 pt, conference]{ieeeconf}  
\usepackage[utf8]{inputenc}
\usepackage[T1]{fontenc}

\IEEEoverridecommandlockouts                              

\usepackage[pdftex]{graphicx} 
\usepackage{amsfonts}
\usepackage[cmex10]{amsmath} 
\usepackage{amssymb}  
\usepackage[dvipsnames,table]{xcolor}
\usepackage{cleveref}
\usepackage[noadjust]{cite}
\usepackage{booktabs}
\usepackage{siunitx}
\usepackage{multirow}
\usepackage{xfrac}
\usepackage{algpseudocode,algorithm,algorithmicx}
\usepackage[caption=false]{subfig}

\crefname{figure}{Fig.}{Figs.}
\Crefname{figure}{Fig.}{Figs.}
\crefname{table}{Table}{Tables}
\Crefname{table}{Table}{Tables}
\newcommand{\train}{\textsc{train}}
\newcommand{\eval}{\textsc{eval}}
\newcommand{\test}{\textsc{test}}
\definecolor{lightlightgray}{gray}{0.9}
\newcommand{\reals}{\mathop{\mathbf{R}}}

\title{\LARGE \bf
Towards a Flexible Deep Learning Method for Automatic Detection of Clinically Relevant Multi-Modal Events in the Polysomnogram
}

\author{Alexander Neergaard Olesen$^{\dagger,1,2,3}$,~\textit{Member, IEEE}, Stanislas Chambon$^{4,5}$, Valentin Thorey$^{5}$, \\Poul Jennum$^{3}$, Emmanuel Mignot$^{2}$ and Helge B. D. Sorensen$^{3}$, \textit{Senior Member, IEEE}
\thanks{Research supported by the Klarman Family Foundation, Technical University of Denmark, and University of Copenhagen with supporting grants from Reinholdt W. Jorck \& Wife's Foundation, Knud Højgaard Foundation, Otto Mønsted Foundation, Vera \& Carl Michaelsens Foundation, Augustinus Foundation, and Stibo Foundation.}
\thanks{$^{\dagger}$Corresponding author: {\tt alexno@stanford / aneol@dtu.dk}}%
\thanks{$^{1}$Department of Health Technology, Technical University of Denmark, 2800 Kgs. Lyngby, Denmark.}%
\thanks{$^{2}$Center for Sleep Sciences and Medicine, Stanford University, Palo Alto, CA 94304, USA.}%
\thanks{$^{3}$Danish Center for Sleep Medicine, University Hospital Copenhagen, 2600 Glostrup, Denmark}%
\thanks{$^{4}$LTCI Telecom ParisTech, Universite Paris-Saclay, Paris, France.}%
\thanks{$^{5}$Research \& Algorithms Team, Dreem, Paris, France.}%
}

\begin{document}

\bstctlcite{IEEEexample:BSTcontrol}

\maketitle
\thispagestyle{empty}
\pagestyle{empty}

\begin{abstract}
Much attention has been given to automatic sleep staging algorithms in past years, but the detection of discrete events in sleep studies is also crucial for precise characterization of sleep patterns and possible diagnosis of sleep disorders. We propose here a deep learning model for automatic detection and annotation of arousals and leg movements. Both of these are commonly seen during normal sleep, while an excessive amount of either is linked to disrupted sleep patterns, excessive daytime sleepiness impacting quality of life, and various sleep disorders. Our model was trained on 1,485 subjects and tested on 1,000 separate recordings of sleep. We tested two different experimental setups and found optimal arousal detection was attained by including a recurrent neural network module in our default model with a dynamic default event window (F1 = 0.75), while optimal leg movement detection was attained using a static event window (F1 = 0.65). Our work show promise while still allowing for improvements. Specifically, future research will explore the proposed model as a general-purpose sleep analysis model.
\end{abstract}

\section{INTRODUCTION}

Analysis of sleep patterns is performed manually by experts in sleep clinics using rules and guidelines defined by the American Academy of Sleep Medicine recently updated in 2018~\cite{Berry2018}. These guidelines outline technical and clinical best practices when performing routine polysomnography (PSG), which is an overnight recording of electroencephalography (EEG), electrooculography (EOG), electromyography (EMG) electrocardiography (ECG), respiratory effort and peripheral limb activity. Expert technicians and somnologists use these physiological variables to analyse sleep patterns and diagnose sleep disorders based on key metrics and indices, such as total sleep time, amount of sleep spent in various sleep stages, and the observed number of discrete events per hour of sleep. Specifically, the number of arousals (short awakenings during sleep, $<$15 s), non-periodic and periodic leg movements (PLM), and the number of apnea events per hour of sleep are summarized in the arousal index (AI), periodic leg movements index (PLMI) and apnea/hypopnea index (AHI), the latter of which is a combination of apneic (no/obstructed respiratory effort) and hypopneic (reduced respiratory effort) events. Excessive amounts of these events are disruptive to normal sleep, which can lead to patient complaints of excessive daytime sleepiness~\cite{Halasz2004}, which in turn is linked to an increase in e.g. automotive accidents and reduced quality of life~\cite{Findley1988}. Increased number of PLMs is also linked to other sleep disorders such as restless legs syndrome, and periodic leg movement disorder~\cite{Ferri2017,AmericanAcademyofSleepMedicine2014}.

Correct diagnosis of sleep disorders is predicated on precise scoring of sleep stages as well as accurate scoring of these discrete sleep events. However, the current gold standard of manual analysis by experienced technicians is inherently biased and inconsistent.Several studies have shown low inter-rater reliability on both the scoring of sleep stages~\cite{Norman2000,Rosenberg2013,Younes2016}, arousals~\cite{Bonnet2007}, and respiratory events~\cite{Rosenberg2014}. Furthermore, manual analysis of PSGs is time-consuming and prone to scorer fatigue. Thus, there is a need for efficient systems that provide deterministic and reliable scorings of sleep studies.

Several recent studies have already explored automatic classification of sleep stages in large cohorts with good results~\cite{Olesen2018,Stephansen2018,Chambon2018a,Biswal2018, Phan2018}, however, the reliable and consistent detection and classification of discrete PSG events in large cohorts remain largely unexplored. 

Recent studies on certain microevents in sleep have indicated that sleep spindles and K-complexes can be reliably detected and annotated with start time and duration using deep learning methods~\cite{Chambon2018b,Chambon2019}. Specifically, these studies proposed a single-shot event detection algorithm, that parallels the YOLO and SSD algorithms used for object detection in 2D images~\cite{Redmon2016,Liu2016}, however, they were limited in scope by detecting events only at the EEG level, and did not explicitly take advantage of the temporal connection of the detected events. Additionally, experiments were carried out on a small-scale database~\cite{Chambon2018b}. 

In this study, we focused on the detection of arousals (AR) and leg movements (LM). These events arise from highly distinct physiological sources, EEG and leg EMG, while ARs are also visible in the EOG and chin EMG. These events are important for the precise characterization of sleep patterns and possible diagnosis of sleep disorders, and an accurate detection is therefore of high interest. We extend previous work in~\cite{Chambon2018b,Chambon2019} by 1) preprocessing and analysing multiple input signals at the same time, and 2) taking into account important temporal context using recurrent neural networks. Furthermore, we apply our model on a larger database than previous studies.

\section{DATA}

\subsection{MrOS Sleep Study}
The MrOS Sleep Study is a part of the larger Osteoporotic Fractures in Men Study with the objective of researching the links between sleep disorders, fractures, cardiovascular disease and mortality in older males ($>65$ years)~\cite{Blank2005,Orwoll2005,Blackwell2011}. Between 2003 and 2005, 3,135 of the original 5,994 participants were recruited to undergo full-night PSG recording at six centers in the US at two separate visits (visit 1 and visit 2) with following 3 to 5-day actigraphy studies at home. The resulting PSG studies were subsequently scored by experienced sleep technicians for standard sleep variables including sleep stages, leg movements, arousals, and respiratory events.

\subsection{Included events and signals}
In this study, we only considered the detection of two PSG events, arousals and leg movements. These events are characterized by a start time and a duration, which we extracted from 2,907 PSG studies from visit 1 available from the National Sleep Research Resource repository~\cite{Dean2016,Zhang2018}. From each PSG study, we extracted left and right central EEG, left and right EOG, chin EMG, and EMG from the left and right anterior tibialis. EEG and EOG channels were referenced to the contralateral mastoid process, while a leg EMG channel was synthesized by referencing left to right. Any PSG without the full set of channels or without any event scoring was eliminated from further analysis.

\subsection{Subset demographics and partitioning}
In total, 2,650 out of the 2,907 PSGs available from visit 1 were included in this study. These were partitioned into \train{}, \eval{}, and \test{} sets of sizes 1,485, 165, and 1,000 studies, respectively. A subset of key demographic and PSG variables are presented in~\cref{tab:demographics}.

\begin{table}[tb]
  \centering
  \footnotesize
  \renewcommand{\arraystretch}{1.3}
  \caption{MrOS subset demographics. Significant \textit{p}-values at $\alpha=0.05$ are shown in bold.}
  \label{tab:demographics}
  \setlength\tabcolsep{5pt}
  \begin{tabular}{lcccc}
    \toprule
                                           & \train{} & \eval{} & \test{} & \textit{p}-value \\
    \midrule
    N                                      & 1,485 & 165  & 1000 & - \\
    Age (years)                            & $76.4 \pm 5.5 $ & $ 76.6 \pm 4.9 $ & $76.4 \pm 5.6 $ & 0.631 \\
    BMI (\si{\kilogram\per\square\second}) & $27.2 \pm 3.8 $ & $ 27.2 \pm 3.4 $ & $27.1 \pm 3.7 $ & 0.879 \\
    AHI (\si{\per\hour})                   & $12.8 \pm 12.9 $ & $ 10.6 \pm 11.8 $ & $11.9 \pm 12.8 $ &\textbf{0.029} \\
    AI (\si{\per\hour})                    & $23.6 \pm 11.5 $ & $ 24.1 \pm 12.2 $ & $23.4 \pm 11.8 $ & 0.607 \\
    PLMI (\si{\per\hour})                  & $34.8 \pm 37.0 $ & $ 37.8 \pm 38.9 $ & $37.3 \pm 38.0 $ & 0.204 \\
    \bottomrule
  \end{tabular}
\end{table}

\section{METHODS}
\begin{figure}
    \centering
    \includegraphics[width=\columnwidth]{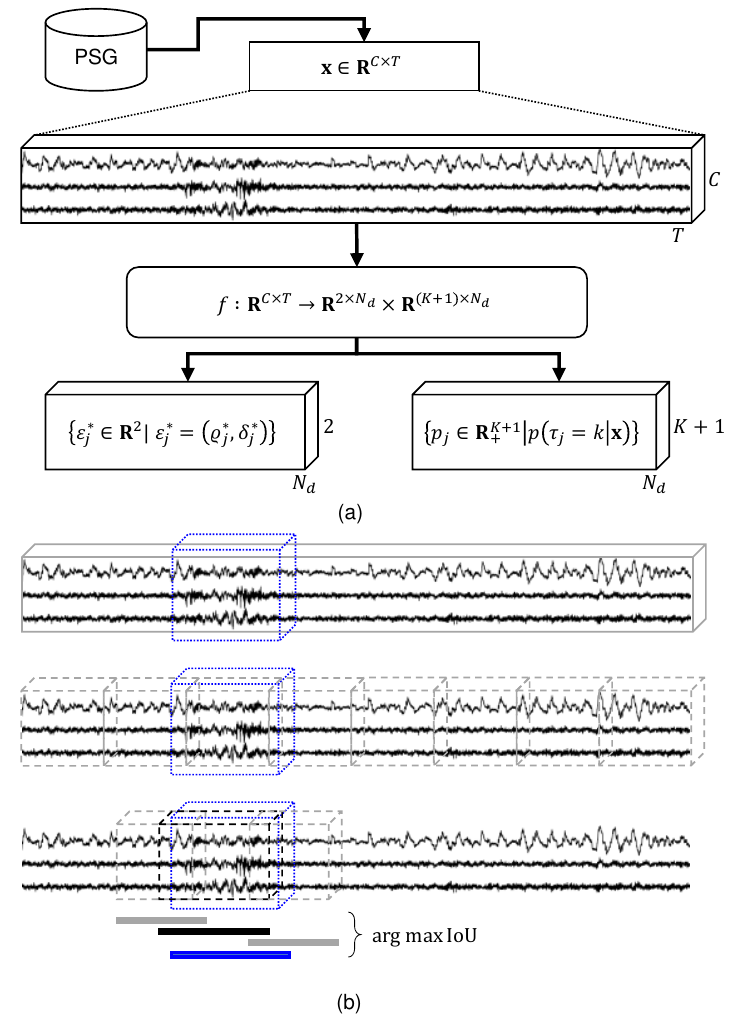}
    
    \caption{Schematic of proposed event detection procedure. (a) Input data $\mathbf{x}$ is fed to the model $f$, which outputs predictions for event classes and localizations for each default event in $\varepsilon^{d}$. (b) The IoU for each predicted $\varepsilon^{*}_{j}$ is then calculated with respect to the true event $\varepsilon_{i}$ and non-maximum suppression is applied to match up true events and predictions. In the current case, the predicted event marked in black has the highest IoU with the true event in blue. For more information, see~\cite{Chambon2018b,Liu2016}.}
    \label{fig:schematic}
\end{figure}
\begin{table*}[t]
  \renewcommand{\arraystretch}{1.3}
  \centering
  \caption{Proposed network architecture. $\phi_C$, linear mixing module; $\phi_{T}$, temporal feature extraction module; $\phi_{R}$, recurrent neural network module; $\psi_{\mathrm{clf}}$, event classification module; $\psi_{\mathrm{loc}}$, event localization module; \textit{C}, number of input channels; \textit{T}, number of samples in segments; $\tilde{C}=2^{2+n_{\max}}$, number of output channels; \textit{K}, number of event classes; $N_{d}$, number of default events in segment; $\tilde{T}=\sfrac{T}{2^{n_{\max}}}$, reduced temporal dimension; bGRU, bidirection gated recurrent unit; ReLU, rectified linear unit.}
  \label{tab:network}
  \begin{tabular}{lcccccccc} \toprule
    Module                                                     & Input dim.                                 & Output dim.                                    & Type           & Kernel size & No. kernels                  & Stride      & Activation \\ \midrule
    $\phi_{C}$                                                 & $\left(C, T\right)$                        & $\left(C, T\right)$                            & 1D convolution & $C$         & $C$                          & 1           & linear \\ \midrule
    \multirow{3}{*}{$\phi_{T,\mathrm{init}}$}                  & $\left(C, T\right)$                        & $\left(8, T\right)$                            & 1D convolution & $3$         & 8                            & $1$         & -- \\
                                                               & $\left(8, T\right)$                        & $\left(8, T\right)$                            & Batch norm.    & --          & 8                            & --          & ReLU \\
                                                               & $\left(8, T\right)$                        & $\left(8, \sfrac{T}{2}\right)$                 & 1D max. pool.  & $2$         & --                           & $2$         & -- \\
    \multirow{3}{2.1cm}{$\phi_{T,k}$ \\ $n=2,\ldots,n_{\max}$} & $\left(2^{n+1}, \sfrac{T}{2^{n-1}}\right)$ & $\left(2^{n+2}, \sfrac{T}{2^{n-1}}\right)$     & 1D convolution & $3$         & $2^{n+2}$                    & $1$         & -- \\
                                                               & $\left(2^{n+2}, \sfrac{T}{2^{n-1}}\right)$ & $\left(2^{n+2}, \sfrac{T}{2^{n-1}}\right)$     & Batch norm.    & --          & $2^{n+2}$                    & --          & ReLU \\
                                                               & $\left(2^{n+2}, \sfrac{T}{2^{n-1}}\right)$ & $\left(2^{n+2}, \sfrac{T}{2^{n}}\right)$       & 1D max. pool.  & $2$         & --                           & $2$         & -- \\ 
    \rowcolor{lightlightgray} $\phi_{R}$ & $(\tilde{C}, \tilde{T})$ & $(2\times \tilde{C}, \tilde{T})$ & bGRU & $\tilde{C}$ & -- & -- & -- & \\ \midrule
    $\psi_{\mathrm{clf}}$                                      & $(\tilde{C}, \tilde{T})$                   & $\left( \left( K + 1 \right) N_{d}, 1 \right)$ & 1D convolution & $\tilde{T}$ & $\left( K + 1 \right) N_{d}$ & $\tilde{T}$ & \parbox[t]{2cm}{softmax for each $K+1$ kernel} \\
    $\psi_{\mathrm{loc}}$                                      & $(\tilde{C}, \tilde{T})$                   & $\left( 2 N_{d}, 1 \right)$                    & 1D convolution & $\tilde{T}$ & $2 N$                        & $\tilde{T}$ & linear \\ \bottomrule
  \end{tabular}
\end{table*}
\subsection{Signal preprocessing}
All signals were resampled to $f_s = \SI{128}{\hertz}$ using poly-phase filtering with a Kaiser window ($\beta = 5.0$) before subsequent filtering according to AASM criteria. Briefly, EEG and EOG channels were subjected to a 4th order Butterworth band pass filter with cutoff frequencies $\left[ 0.3, 35.0 \right]$ Hz, while chin and leg EMG channels were filtered with a 4th order Butterworth high pass filter with a 10 Hz cutoff frequency. All filters employed zero-phase filtering. Lastly, each channel was normalized by subtracting the channel mean and dividing by the channel standard deviation across the entire night.


\subsection{Detection model overview}
In brief, the proposed model receives as input a tensor $\mathbf{x}\in \mathbf{R}^{C \times T}$ containing $C$ channels of data in a segment of $T$ samples, along with a set of events $\lbrace \varepsilon_{i} \in \reals^{2} \mid \varepsilon_i = \left( \varrho_{i}, \delta_{i} \right),\, i=1,\ldots,N_{\mathbf{x}} \rbrace$, were $N_{\mathbf{x}}$ is the number of events in the associated time segment and $\left(\varrho_{i},\delta_{i}\right)$ are the start time and duration of event $\varepsilon_i$. The objective of the deep learning model $f$ is then to infer $\lbrace \varepsilon_{i} \rbrace$ given $\mathbf{x}$. To do this, a set of default events $\lbrace \varepsilon_{j}^{d} \in \reals^2 \mid j=1,\ldots,N_{d},\, N_{d} = T/\tau \rbrace$ is generated over the segment of $T$ samples, where $\tau$ is the size of each default event window in samples. The model outputs probabilities for $K$ classes including the default, non-event class for each default event window. The probability for a given class $k$ in the default event window $\varepsilon_j^d$ must be greater than a classification threshold $\theta_{\mathrm{clf}}$. In order to select among many possible candidates of predicted events, all predicted events of class $k$ over the possible events in $N_d$ is subjected to non-maximum suppression using the intersection-over-union (IoU, Jaccard index) as in~\cite{Redmon2016}. A high-level schematic of the detection model is shown in~\cref{fig:schematic}.


\subsection{Network architecture}
The architecture for the proposed PSG event detection model follows closely the event detection algorithms described in~\cite{Chambon2018b,Chambon2019}, albeit with some specific changes. An overview of the proposed network in the model $f$ is provided in~\cref{tab:network}. Briefly, the model comprises three modules:
\begin{enumerate}
\item a channel mixing module $\phi_{C} : \mathbf{R}^{C \times T} \to \mathbf{R}^{C \times T}$;
\item a feature extraction module $\phi_{T} : \mathbf{R}^{C \times T} \to \mathbf{R}^{\tilde{C} \times \tilde{T}}$;
\item and an event detection module $\psi$,
\end{enumerate}
the latter containing two submodules performing event classification $\psi_{\mathrm{clf}} : \mathbf{R}^{\tilde{C} \times \tilde{T}} \to \mathbf{R}^{(K+1)\times N_{d}} $ and event localization $\psi_{\mathrm{loc}} : \mathbf{R}^{\tilde{C} \times \tilde{T}} \to \mathbf{R}^{2 \times N_{d}}$, respectively. The difference between these two submodules is that $\phi_{\mathrm{clf}}$ outputs the probability of the default, non-event class and $K$ event classes, while $\phi_{\mathrm{loc}}$ predicts a start time and a duration of all predicted events relative to a specific default event window.
The channel mixing module $\phi_{C}$ receives a segment of input data $x \in \mathbf{R}^{C \times T}$, where $C$ is the number of input channels and $T$ is the number of time samples in the given segment, and subsequently performs linear channel mixing using 1D~convolutions to synthesize $C$ new channels. 
Following $\phi_{C}$, the feature extraction module $\phi_{T}$ consists of $n_{\max}$ blocks with the first block $\phi_{T,1} : \mathbf{R}^{C \times T} \to \mathbf{R}^{8 \times \sfrac{T}{2}}$ and the $n$th block $\phi_{T,n} : \mathbf{R}^{2^{n+1} \times \sfrac{T}{2^{n-1}}} \to \mathbf{R}^{2^{k+2} \times \sfrac{T}{2^{n}}}$. All $n_{\max}$ blocks implement $\phi_{T,n}$ using 1D~convolution layers followed by batch normalization of the feature maps, rectified linear unit activation, and final 1D maximum pooling layers across the temporal dimension. Kernel sizes and strides for convolution and max. pool. layers in $\phi_{T}$ were set to 3 and 1, and 2 and 2, respectively, while the number of feature maps in $\phi_{T,n}$ was set to $2^{n+2}$. The event classification submodule $\psi_{\mathrm{clf}}$ is implemented a 1D convolution layer across the entire data volume using $(K+1)N_{d}$ feature maps of size and stride $\tilde{T} = T/2^{n_{\max}}$, where $K \in \mathbf{N}$ is the number of event classes to be detected and $N_{d} \in \mathbf{N}$ is the number of default event windows. The event localization submodule $\psi_{\mathrm{loc}}$ is likewise implemented using a 1D convolution layer across the entire data volume.

\subsection{Data and event sampling}
The proposed network requires an input tensor $x \in \mathbf{R}^{C \times T}$ containing PSG data in the time segment of size $T$ as well as information about the associated events in the segment. Since the total number of segments in a standard PSG without any event data far outnumbers the number of segments with event data, we implemented a random sampling of non-event and event classes with the sampling probability of class $k$ inversely proportional to the number of classes, such that $p_k = \frac{1}{K+1},\,k=\left[0\,..\,K \right]$, where $k=0$ is the default (non-event) class. At training step $t$, we thus sample a class $k$ and afterwards randomly sample a single class $k$ event $\varepsilon_{k}$ between all class $k$ events. Finally, we extract a segment of PSG data of size $C \times T$ with start of segment in the interval $\left[ \bar{\varepsilon}_{k} - T, \bar{\varepsilon}_{k} + T \right]$, where $\bar{\varepsilon}_{k}$ is the sample midpoint of $\varepsilon_{k}$. This ensures that each $\mathbf{x}$ overlaps 50\% with at least one associated event.

\subsection{Optimization of network parameters}
The network parameters were optimized using mini-batch stochastic gradient descent with initial learning rate of $10^{-3}$ and a momentum of $0.9$. Minibatches were balanced with respect to the detected classes. The optimization was performed with respect to the same loss function described in~\cite{Chambon2018b,Chambon2019} and the network was trained until convergence determined by no decrease in the loss on the \eval{} set over 10 epochs of \train{} data. We also employed learning rate decay with a factor of 2 every 5 epochs of non-decreasing \eval{} loss.

\subsection{Experimental setups}
In this study, we examined two different experimental setups. 
\paragraph{Experiment A} First, we investigated the differences in predictive performance using a static vs. a dynamic default event window size. This was realized by running six separate training runs with $\tau \in \lbrace 3, 5, 10, 15, 20, 30 \rbrace \times f_{s}$, as well as a single training run where $f$ was evaluated for all $\lbrace 3, 5, 10, 15, 20, 30 \rbrace \times f_{s}$. The best performing model was determined by evaluating F1 score on the \eval{} set for both LM and AR detection. 
\paragraph{Experiment B} Second, we tested a network where we added a recurrent processing block $\phi_{R}$ after the feature extraction block $\phi_{T}$ as shown in grey in~\cref{tab:network}. We considered a single bidirectional gated recurrent unit (bGRU) layer with $\tilde{C}$ units. Predictions were evaulated across multiple time-scales $\tau \in \lbrace 3, 5, 10, 15 \rbrace \times f_{s}$ 

All experiments were implemented in PyTorch 1.0~\cite{Paszke2017}.

\subsection{Performance metrics}
All models were evaluated on the \eval{} and \test{} sets using precision (Pr), recall (Re), and F1 scores (F1):
\begin{align*}
    \mathrm{Pr} &= \frac{\mathrm{TP}}{\mathrm{TP} + \mathrm{FP}}, \quad \mathrm{Re} = \frac{\mathrm{TP}}{\mathrm{TP} + \mathrm{FN}} \\
    \mathrm{F1} &= 2 \frac{\mathrm{Pr} * \mathrm{Re}}{\mathrm{Pr} + \mathrm{Re}} = \frac{2\mathrm{TP}}{2\mathrm{TP} + \mathrm{FP} + \mathrm{FN}},
\end{align*}
where TP, FP, and FN, are the number of true positives, false positives and false negatives, respectively.

\subsection{Statistical analysis}
Demographic and polysomnographic variables were tested for subset differences with Kruskall-Wallis H-test for independent samples.

\section{RESULTS AND DISCUSSION}

    

\begin{figure}
    \centering
    \subfloat[]{%
    	\includegraphics[width=\columnwidth]{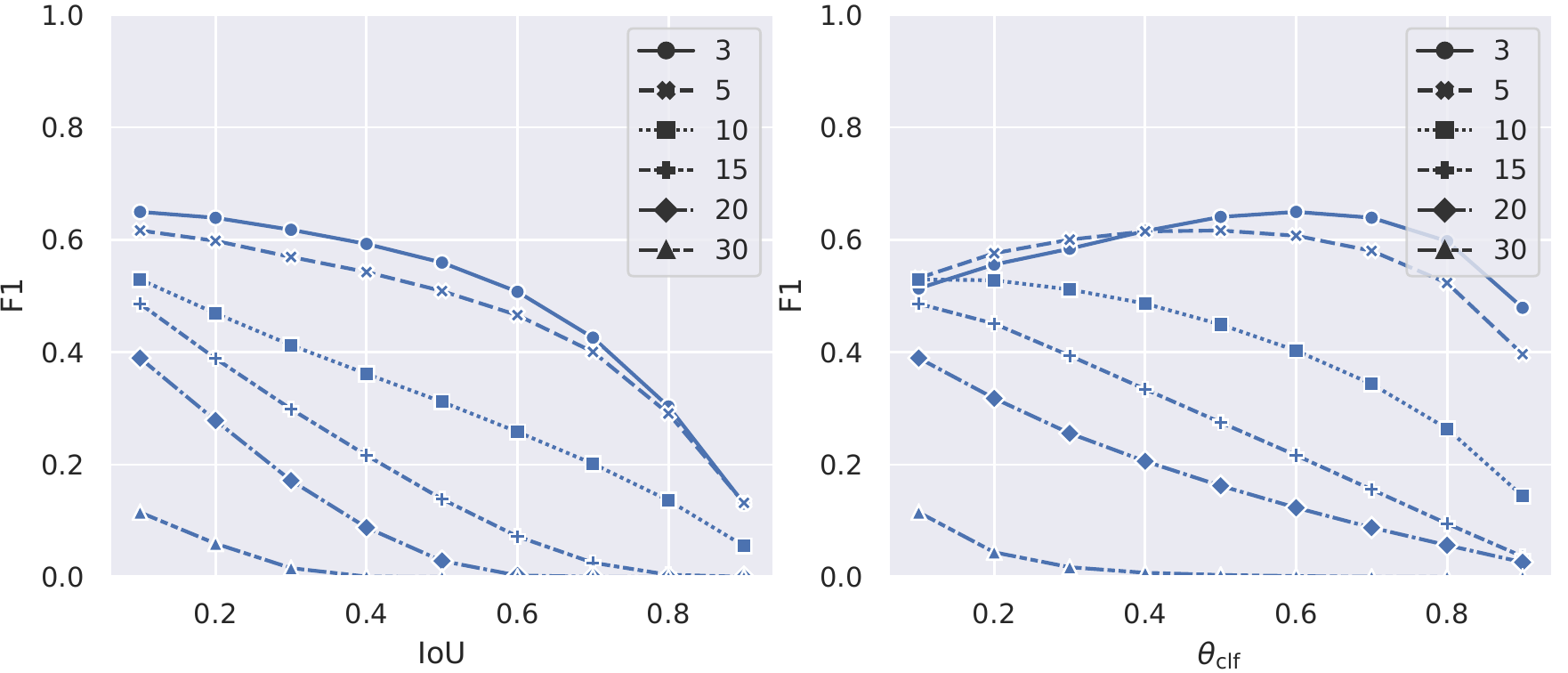}%
    	\vspace{-0.75cm}%
        \label{fig:lm_static}%
    }
    \par\medskip
    \subfloat[]{%
    	\includegraphics[width=\columnwidth]{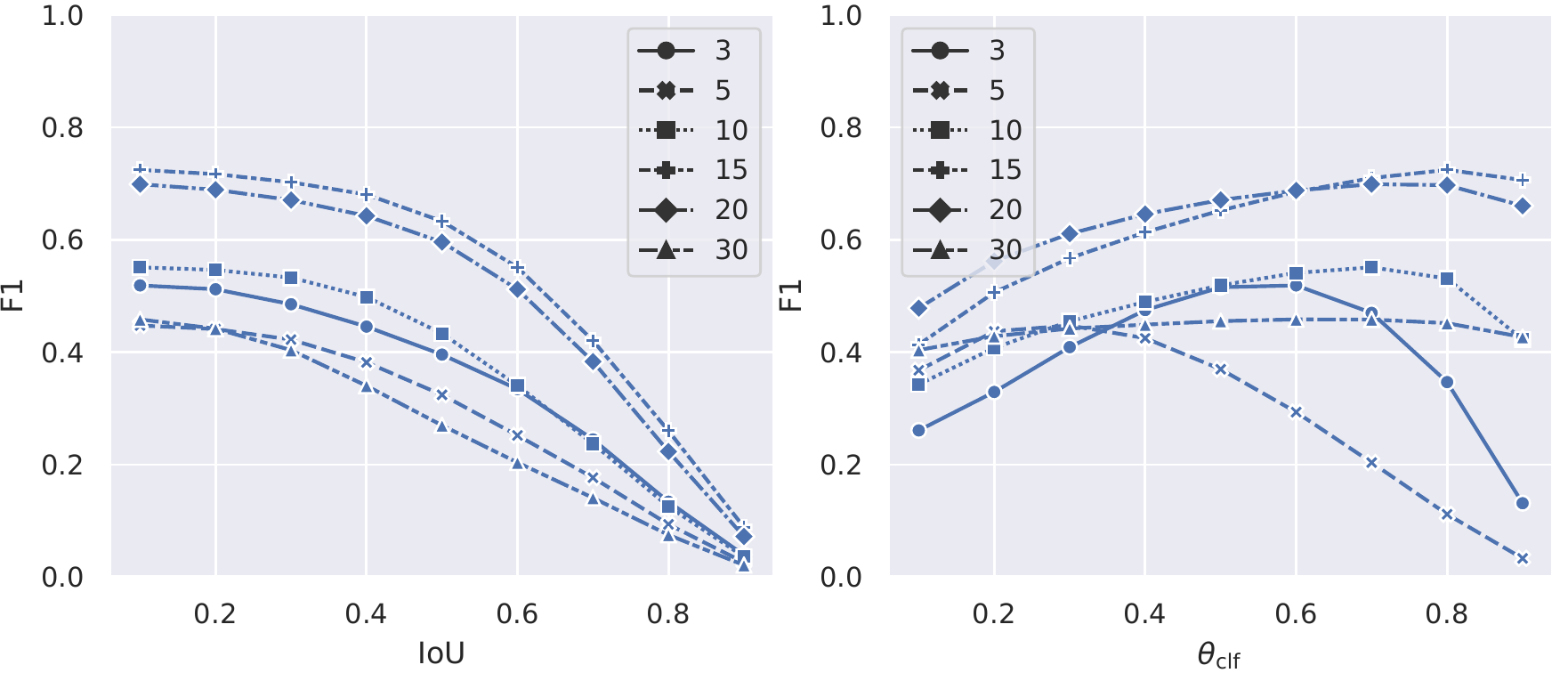}%
    	\vspace{-0.75cm}%
        \label{fig:ar_static}%
    }
    \par\medskip
    \subfloat[]{%
    	\includegraphics[width=\columnwidth]{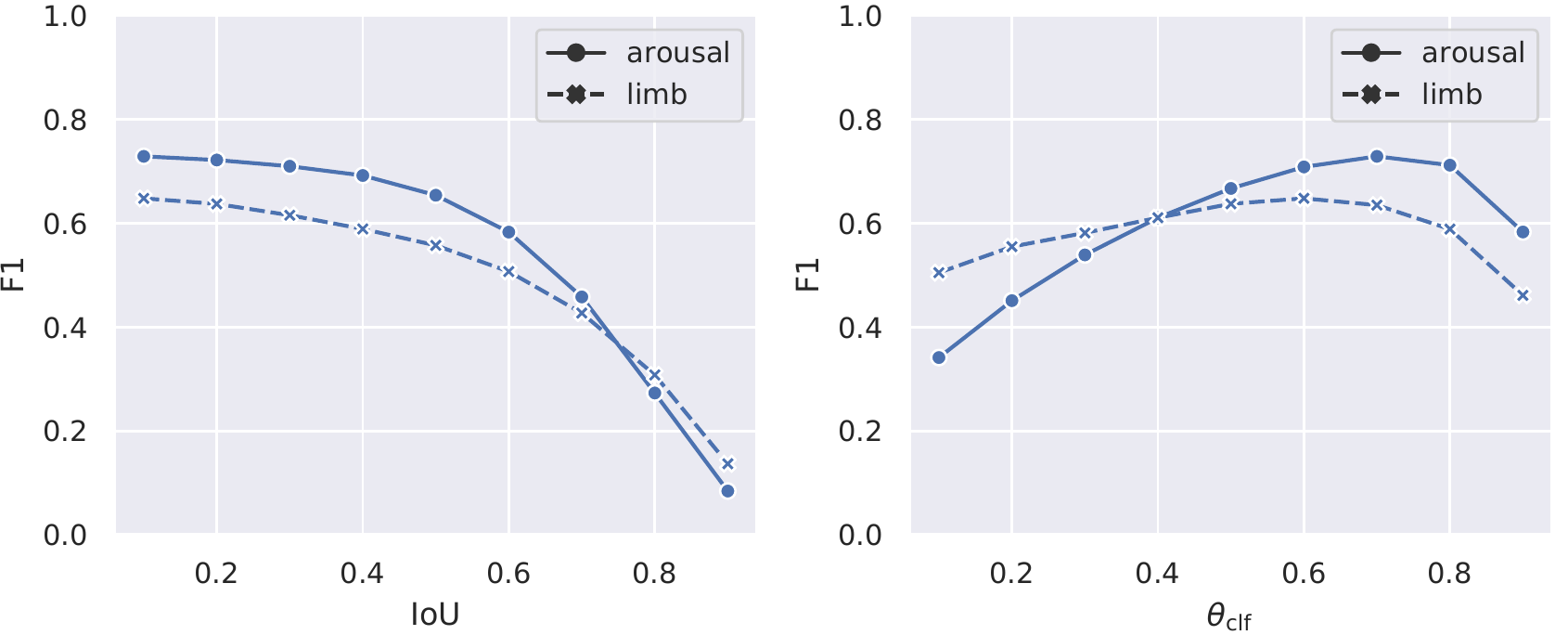}%
    	\vspace{-0.75cm}%
        \label{fig:ar_lm_dynamic}%
    }
    \caption{Experiment A: Optimizing IoU and $\theta_{\mathrm{clf}}$ in static models on the \eval{} set by varying default event window size in seconds in $\lbrace3, 5, 10, 15, 20, 30 \rbrace$ (a)-(b). Left panels show the IoU vs. F1 score, while right panels show classification threshold $\theta_{\mathrm{clf}}$ against F1 score. (a) LM model. Here, the model performs best for $\mathrm{IoU}=0.1$ and $\theta_{\mathrm{clf}} = 0.6$ using a window size of $\tau=\SI{3}{\second}\times f_{s}$. (b) AR model. Here, the model performs best for $\mathrm{IoU}=0.1$ and $\theta_{\mathrm{clf}} = 0.8$ using a window size of $\tau=\SI{15}{\second} \times f_{s}$. (c) Dynamic models show optimal performance for $\mathrm{IoU}=0.1$ and $\theta_{\mathrm{clf}}=0.7$ and $\theta_{\mathrm{clf}}=0.6$ for AR and LM detection, respectively.}
    \label{fig:experiment_a}
\end{figure}

Shown in~\cref{fig:lm_static,fig:ar_static} are the F1 scores as a function of IoU and the classification threshold $\theta_{\mathrm{clf}}$ for both the LM and AR detection models. It is apparent that both models perform best with a minimum overlap ($\mathrm{IoU} = 0.1$) with their respective annotated events, and do not benefit from increasing the overlap. This might be caused by the fact that the annotated events might not be precise enough, and not due to issues with the model itself. For example, it is not uncommon to only mark the beginning of an event in standard sleep scoring software, as the duration will automatically be annotated by a default length, such as 3 s for ARs, and 0.5 s for LM (which is the minimum duration as defined by the AASM guidelines~\cite{Berry2018}). Future studies will be able to confirm this by either collecting a precisely annotated cohort, or by investigating the average start time and duration discrepancies between annotated and predicted events. 

It is also apparent from~\cref{fig:lm_static,fig:ar_static} that both detection models benefit from imposing a strict classification threshold. Specifically, LM detection performance as measured by F1 was highest with $\theta_{\mathrm{clf}} = 0.6$, while maximum AR detection performance was attained with an even higher $\theta_{\mathrm{clf}}$ of 0.8.

Furthermore, we explored allowing for multiple time-scales in the dynamic models, shown in~\cref{fig:ar_lm_dynamic}. It was hypothesized that having the default event windows dynamic instead of static would allow for more flexibility and thus better predictive performance, however, we observed no significant differences between the optimal static window and the dynamic window model.

\begin{figure}
    \centering
    \includegraphics[width=\columnwidth]{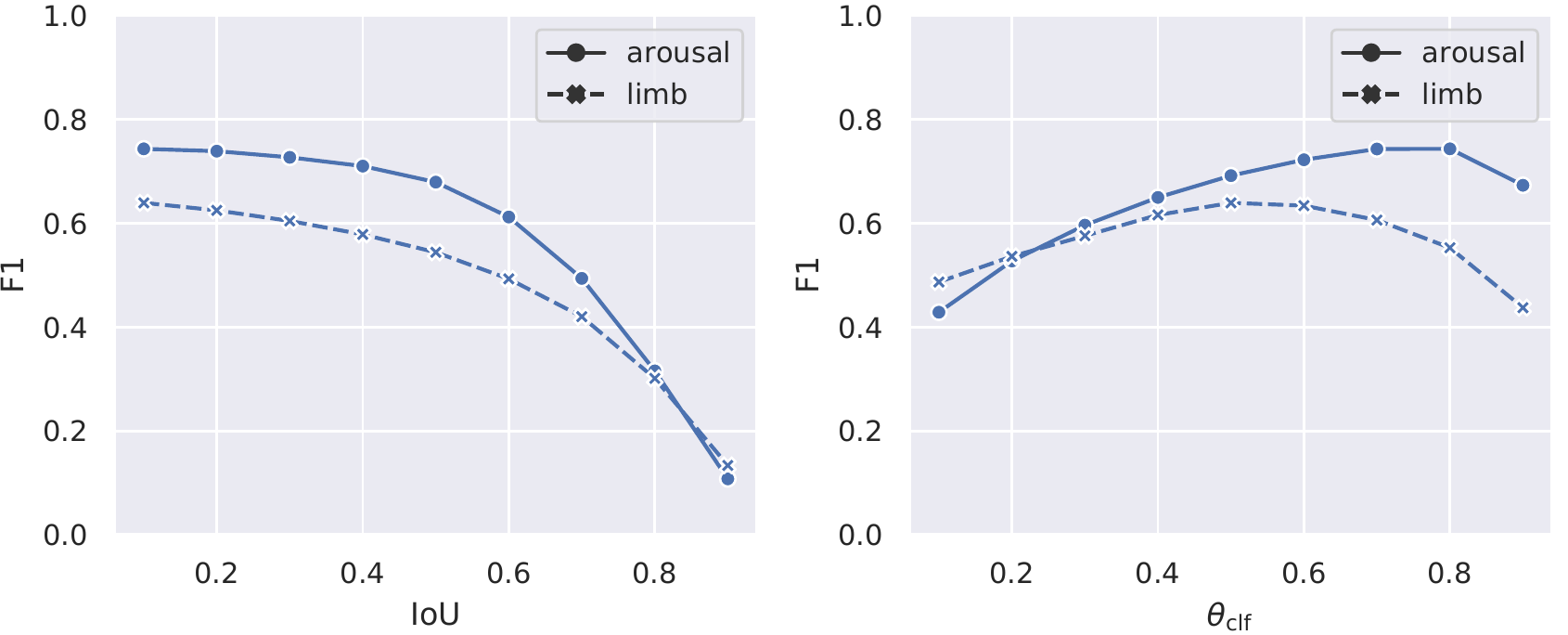}
    \caption{Experiment B. F1 performance on the \eval{} set as a function of IoU and $\theta_{\mathrm{clf}}$ for AR and LM detection when adding the $\phi_{R}$ module. Best performance is seen for $\mathrm{IoU}=0.1$ for both AR and LM detection, and $\theta_{\mathrm{clf}} = 0.6$ and $\theta_{\mathrm{clf}}=0.8$ for LM and AR detection, respectively.}
    \label{fig:experiment_b}
\end{figure}

Shown in~\cref{fig:experiment_b} are the performance curves for the RNN (bidirectional GRU) version of the proposed model for each of the two event detection tasks. While the optimal IoU and $\theta_{\mathrm{clf}}$ points are unchanged from the static/dynamic models presented in~\cref{fig:experiment_a}, the optimal F1 value for AR detection is increased by incorporating temporal dependencies in the model. The reverse is true for LM detection, which saw a slight decrease in predictive performance caused by a lower precision (see~\cref{tab:test_results}). Future work should consider optimizing predictive performance by investigating the effects of varying the number of bGRU layers and the number of hidden units in $\phi_{R}$, since this was not performed here. 

\begin{figure}[tb]
    \centering
    \includegraphics[width=\columnwidth]{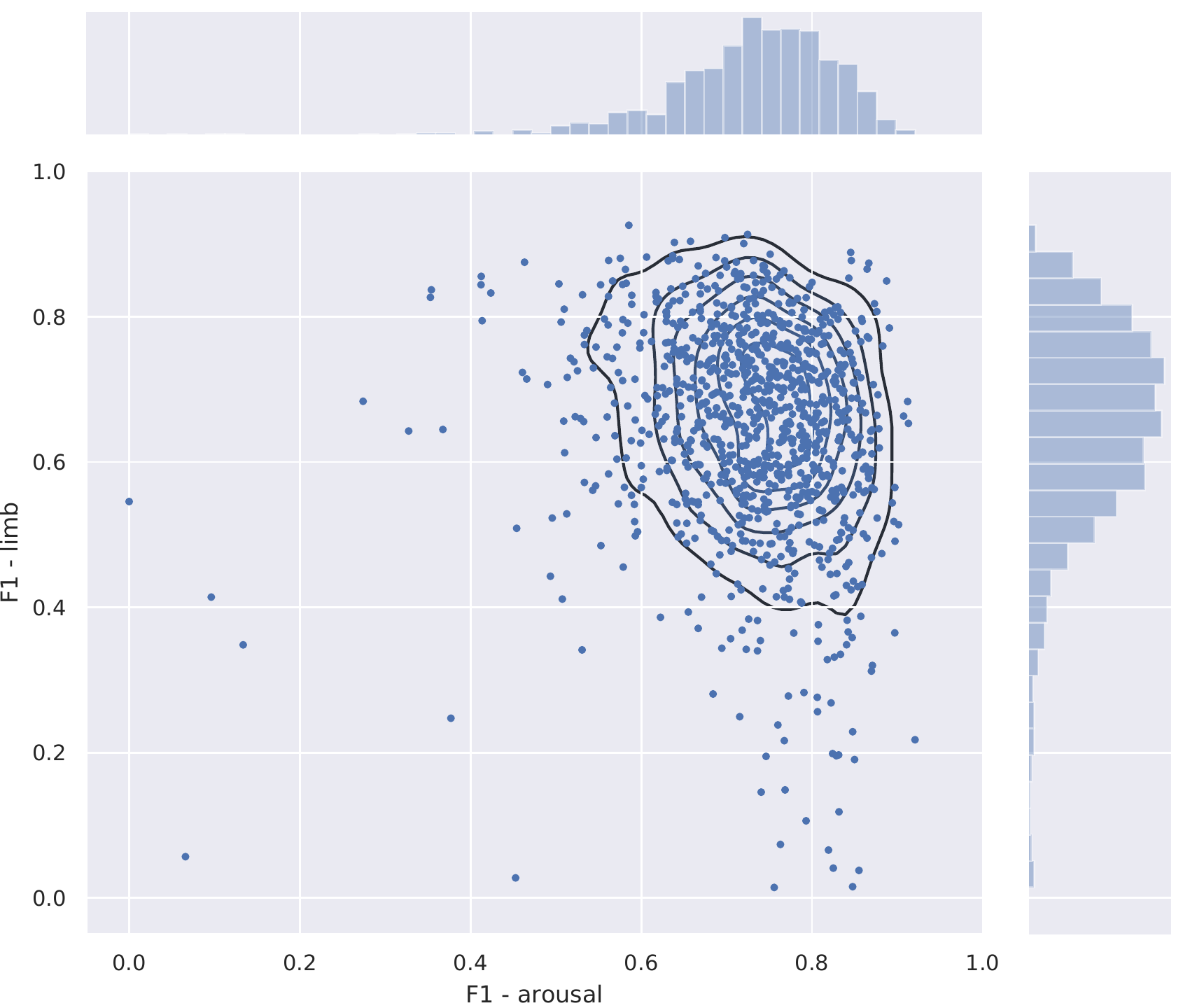}
    \caption{Visualization of F1 scores for both AR and LM detection using the dynamic model.}
    \label{fig:test_distribution}
\end{figure}

Application of the optimal models on the \test{} data is shown in~\cref{tab:test_results}. We observed that with the given architecture of $f$ and the given labels and input data in \train{}, LM detection was maximal for the model with a static/dynamic window, while adding a recurrent module only positively impacted AR prediction. We observed a general decrease in both precision and recall for LM detection when adding $\phi_{R}$, while precision actually increased and recall decreased for AR detection. An example visualization of the joint distribution of F1 scores obtained from the dynamic model applied to the \test{} data is shown in~\cref{fig:test_distribution}. While some outliers are readily observable especially for LM detection, the majority of subject F1 scores follows an approximate bivariate normal distribution.

Subset partitions were reasonably well-distributed with no significant differences between key variables, see~\cref{tab:demographics}. An exception is the AHI, although the associated effect is small and most likely a result of the low sample size in \eval{} compared to \train{} and \test{}. It is noted, that although AHI, AI, and PLMI are not normally distributed and summarizing these variables with standard deviations is invalid, it is nevertheless standard practice in sleep medicine and thus presented the same way here. We performed little data cleaning in order to provide as much data and variation to the deep learning model as possible, however, future efforts should explore and apply inclusion criteria such as minimal total sleep time, artifact detection and removal of studies with severe artifacts. We did impose a trivial lower bound on the number of scored events (>0) for a PSG to be included in this study, but stricter requirements could potentially improve model performance.

In this work, we investigated 'systemic' PSG events present in multiple signal modalities instead of EEG-specific events, which required changes to the network architecture. Specifically, we kept the signal modality encoded in the first dimension of the tensor propagated through the network, which allowed for the use of one-dimensional convolutional operators. By performing 1D convolutions and keeping the channel information in the feature maps instead of keeping them as separate dimensions and performing 2D convolutions as proposed in~\cite{Chambon2018b,Chambon2019}, we simplify and reduce the number of computations and training time by a factor $\propto C$. 
\begin{table}[tb]
    \centering
    \footnotesize
    \caption{Application of optimized models on \test{} data. Data are shown as subject-averaged F1, precision (Pr) and recall (Re) with associated standard deviations. Top four rows correspond to Experiment A, while bottom two rows correspond to Experiment B. AR: arousal; LM: leg movement; RNN: recurrent neural network.}
    \label{tab:test_results}
    \begin{tabular}{clccc} \toprule
        Model & F1 & Pr & Re \\ \midrule
        LM, static & $0.648 \pm 0.148$ & $0.631 \pm 0.181$ & $0.720 \pm 0.141$ \\
        AR, static & $0.727 \pm 0.102$ & $0.706 \pm 0.113$ & $0.771 \pm 0.132$ \\
        LM, dynamic & $0.647 \pm 0.148$ & $0.627 \pm 0.181$ & $0.722 \pm 0.14$ \\
        AR, dynamic. & $0.729 \pm 0.102$ & $0.699 \pm 0.115$ & $0.785 \pm 0.131$ \\ \midrule
        LM, RNN & $0.639 \pm 0.147$ & $0.606 \pm 0.180$ & $0.727 \pm 0.126$ \\
        AR, RNN & $0.749 \pm 0.105$ & $0.772 \pm 0.107$ & $0.748 \pm 0.138$ \\ \bottomrule 
    \end{tabular}
\end{table}
However, we did not investigate the effects of modeling the conditional probability of AR and LM occurrence, but the proposed architecture is versatile enough to detect both events jointly as well as separately. Previous work also suggest that detecting multiple objects at the same time is of high interest and leads to (at least) non-inferior performances~\cite{Chambon2018b, Chambon2019, Redmon2016, Liu2016}.

Additionally, we speculated that the temporal dynamics of the PSG signals were important for optimal event detection performance. Although the effects were small, we did show an increase in F1 score in AR detection when adding an RNN module to the network before the detection module. However, this was not the case for LM detection, which is most likely due to the different temporal and physiological characteristics of the two events in question. 

Future efforts will be addressing the fact that in the current modeling scheme, events are mutually exclusive given a certain default event window size. However, it is common to see ARs and LMs as a result of one another, and thus, if the window size is too small, a more unlikely event as measured by classification threshold and IoU will be removed even if it matches up to a specific true event of a certain class.

\section{CONCLUSIONS}
We have proposed a deep learning model that extends on previous work and shows promise in automatic detection of arousals and leg movements during sleep. The proposed model is flexible in allowing for the detection of multiple events of distinct physiological natures. Future work will expand further on adding more signals and event classes in order to complete a general purpose sleep analysis tool.


\section*{ACKNOWLEDGMENT}

Some of the computing for this project was performed on the Sherlock cluster. We would like to thank Stanford University and the Stanford Research Computing Center for providing computational resources and support that contributed to these research results.


\bibliographystyle{IEEEtran}
\bibliography{IEEEabrv,library}

\end{document}